\documentclass[12pt]{iopart}
\usepackage{epsfig}
\usepackage{amssymb}
\usepackage{bm}
\begin{document}

\title[Tissue Plasticity and Morphogenesis]{Cell resolved, multiparticle model of plastic tissue deformations and morphogenesis}

\author{ Andras Czirok $^{1,2}$\footnote{corresponding author} ,
	 Dona Greta Isai $^{2}$
       }
\address{$^1$ Department of Anatomy \& Cell Biology, University of Kansas Medical Center, Kansas City, KS, USA}
\address{$^2$ Department of Biological Physics, Eotvos University, Budapest, Hungary}

\ead{aczirok@kumc.edu}

\begin{abstract}

We propose a three dimensional mechanical model of embryonic tissue dynamics. Mechanically coupled adherent cells are represented as particles interconnected with elastic beams which can exert non-central forces and torques. Tissue plasticity is modeled by a stochastic process consisting of a connectivity change (addition or removal of a single link) followed by a complete relaxation to mechanical equilibrium. In particular, we assume that (i) two non-connected, but adjacent particles can form a new link; and (ii) the lifetime of links is reduced by tensile forces.  We demonstrate that the proposed model yields a realistic macroscopic elasto-plastic behavior and we establish how microscopic model parameters affect the material properties at the macroscopic scale. Based on these results, microscopic parameter values can be inferred from tissue thickness, macroscopic elastic modulus and the magnitude and dynamics of intercellular adhesion forces. In addition to their mechanical role, model particles can also act as active simulation agents and modulate their connectivity according to specific rules. As an example, anisotropic link insertion and removal probabilities can give rise to local cell intercalation and large scale convergent extension movements. The proposed stochastic simulation of cell activities yields fluctuating tissue movements which exhibit the same autocorrelation properties as empirical data from avian embryos.

\end{abstract}

\vspace{2pc}

\maketitle

\def\be{\begin{equation}}
\def\ee{\end{equation}}
\def\bea{\begin{eqnarray}}
\def\eea{\end{eqnarray}}
\def\r{\mathbf{r}}
\def\p{\mathbf{p}}
\def\hd{\hat{\mathbf{d}}}
\def\x{\mathbf{x}}
\def\n{\mathbf{n}}
\def\t{\mathbf{t}}
\def\F{\mathbf{F}}
\def\P{\mathbf{P}}
\def\Q{\mathbf{Q}}
\def\M{\mathbf{M}}
\def\N{\mathbf{N}}
\def\R{\mathbf{R}}
\def\V{\mathbf{V}}
\def\v{\mathbf{v}}
\def\u{\mathbf{u}}
\def\bphi{\bm{\phi}}

\section{Introduction}
Tissue engineering, the controlled construction of tissues -- cells and their extracellular matrix (ECM) environment -- is a promising avenue for future biomedical innovations. To realize this possibility, the dynamic and mutually interdependent relationship between cell and tissue movements has to be understood. The cytoskeleton as well as embryonic tissues are dynamic structures, capable of both relaxing and generating mechanical stress. A key, and little explored mechanical component of sustained tissue movement is plastic behavior \cite{Preziosi10, Preziosi11, Ranft10, Giverso12} -- irreversible alteration of the driving force-free tissue shape. Plasticity is clearly important during embryonic development as stresses do not accumulate in embryonic tissues, despite the large deformations. 

Deformation of physical objects subjected to external or internal forces are usually calculated by the partial differential equations (PDE) of continuum mechanics -- this approach uses spatially resolved mechanical stress and strain tensors as model variables \cite{Fung93}. Active biomechanical processes are usually modeled using a spatial decomposition and re-assembly method \cite{Stolarska09}: if the behavior (growth, shape change) of specific parts of the structure are known in mechanical isolation (free boundary conditions), then constraining adjacent parts to form a smooth continuum can yield the deformation of the whole composite structure. In this manner, one can evaluate how autonomous shape changes in one part of the embryo (prescribed ``growth laws'') can drive tissue movements elsewhere \cite{Taber09, Varner10}.  In this approach, the cellular origin of the ``growth laws'' is not explained. Yet, this is a challenging problem as most often the cellular changes are not just the scaled-down reflections of tissue deformations. For example, vertebrate embryos elongate substantially during early development, yet cells of the epiblast maintain their isotropic aspect ratio. Therefore, several tissue-level effects are puzzling outcomes of cellular activities, which are ``purposeful'' changes in cell-cell and cell-ECM contacts. 

To connect cell activity such as active intercalation or collective migration to tissue movements it is often advantageous to model individual cells and obtain the tissue scale behavior through computer simulations. Widely used models for cell-cell interactions represent individual adherent cells as “fluid” droplets, like the cellular Potts model \cite{Graner92, Izaguirre04} and its grid-free version, the subcellular element model \cite{Newman05, Sandersius11a}. These model choices are motivated by the demonstrated non-Newtonian fluid-like behavior of simple cell aggregates \cite{Forgacs98}. Cell-based models have been used to formulate hypotheses for cell activity (such as chemotactic guidance) and used to develop simulations to obtain tissue movements \cite{Zajac00, Vasiev10, Sandersius11c}. Such models are, however, not yet used to predict mechanical stresses. Furthermore, as discussed  recently \cite{Szabo11} these simulations often include biomechanical artifacts such as friction with a non-existing reference frame. In particular, within the freely floating embryonic cell mass the momentum is conserved (while the momentum is not conserved when cells can exert traction on an underlying surface).

Another class of models used to model epithelial morphogenesis is termed ``vertex models'', in which each cell is represented by a polygon corresponding to the cell membrane \cite{Honda80, Fletcher14}. The polygon vertices are usually points belonging to the boundary of three adjacent cells.  Cell movements are due to the motion of these vertices, which in turn are often assumed to be driven by cortical cytoskeletal contractivity, surface tension due to intercellular adhesion and hydrostatic pressure difference between adjacent cells \cite{Farhadifar07}. Such models can also include cell neighbor exchange, perhaps the most significant is the T1 transition in which two adjacent cells are separated by the contact of their immediate lateral neighbors. These models are extremely suitable to describe morphogenetic movements in Drosophila \cite{Rauzi08} and zebrafish \cite{Maitre12}, where the contractile cytoskelton is localized in a thin cortical layer adjacent the cell membrane. While most studies utilizing the vertex models are confined to two dimensions, a generalized vertex model which describes cells as three dimensional prisms was also proposed \cite{Honda04}.

In this paper we introduce a cell-resolved, three dimensional off-lattice model of tissue layers that (i) does not assume that cytoskeletal contractility is localized in the cortical cytoskeleton, and (ii) is computationally simpler than the 3D vertex models. In the model proposed here, one particle represents a single cell, but in contrast with similar cell-center or spheroid models \cite{Drasdo00,RamisConde08}, the mechanical connectivity of the cells are explicitly represented as elastic beams connecting adjacent particles. The beams can be compressed, stretched, bent and twisted. Therefore, instead of central forces, these links can exert torques and forces that are not parallel to the line connecting the particles. We assume that the tissue is always in mechanical equilibrium, i.e., cellular activity (contraction, rearrangement of the links) is slow compared to the time needed for the environment to accommodate these changes. The great advantage of explicit cell-cell contact representation is that we can formulate certain cell activities or plastic stress relaxation as rule sets that specify probabilities for the removal and insertion of links, or alter the equilibrium properties of existing links.  Thus, as we demonstrate, convergent-extension movements can be simulated by preferentially creating new links along one direction while removing links in the perpendicular direction.

\section{Model}
\subsection{Mechanics}
\begin{figure}
\begin{center}
\includegraphics[width=5in]{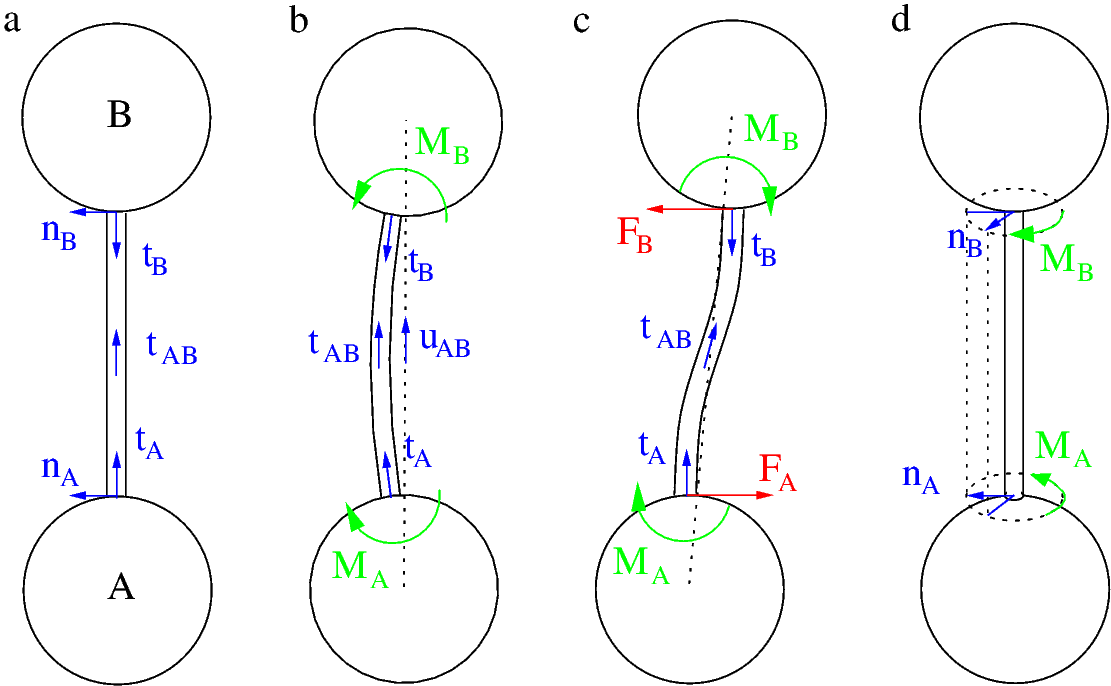}
\caption{\small
The basic mechanical model. a: Two particles, $A$ and $B$ are shown interconnected with a link. At both ends of the link a pair of unit vectors, $\t_A, \n_A$ and $\t_B,\n_B$, specify the link direction and orientation.  The link index is omitted in the figure for better transparency.  The direction and orientation  vectors co-rotate with the particle attached to the link. The direction of the link at its midpoint between particles A and B is denoted by the unit vector $\t_{AB}$. When the link is stress free, $\t_A$, $\t_B$ and $\t_{AB}$, as well as $\n_A$ and $\n_B$ are co-linear. b: A symmetric rotation of both particles yields torques $\M_A$ and $\M_B$ acting on particles $A$ and $B$, respectively. These torque vectors are perpendicular to the plane of the figure. The unit vector pointing from particle $A$ to $B$ is denoted by $\u_{AB}$. In this configuration the link does not exert forces perpendicular to $\u_{AB}$ as the net torque $-(\M_A+\M_B)$ acting on the link is zero. c: A lateral misalignment of the particles creates torques $\M_A$, $\M_B$ and also  shear forces $\F_A$, $\F_B$ acting at the particle-link junctions.  d: Link torsion is characterized by the angle between the unit vectors $\n_A$ and $\n_B$. A twisted link rotates the particles around the link axis to reduce torsion. 
}
\label{fig_model}
\end{center}
\end{figure}

In our model cells are represented as particles, characterized by their position and orientation ($\r_A$ and $\bphi_A$ for particle $A$, respectively). The mechanical connection between cells is modeled with links that can exert non-central forces and torques. As we focus on mechanical equilibrium, the inertia (mass) of these model objects are irrelevant.

Torques are exerted if links are deformed: we envision a behavior similar to that of coil springs. A pair of unit vectors,  $\t_{A,l}$ and $\n_{A,l}$, specify the mechanically neutral link direction and orientation of link $l$ at particle $A$ (Fig.~\ref{fig_model}a). These vectors co-rotate with the particle:
 \be
 \t_{A,l} = \R(\bphi_A) \t_{A,l}^{(0)}
 \label{rot1}
 \ee
and
 \be
 \n_{A,l} = \R(\bphi_A) \n_{A,l}^{(0)}
 \label{rot2}
 \ee
where $\R$ is the rotation operator and $\t_{A,l}^{(0)}$ and $\n_{A,l}^{(0)}$ denote the neutral link directions in the initial configuration where $\bphi_A=0$. 

A link $l$ is bent if its preferred direction at the particle $\t_{A,l}$ is distinct from its actual direction $\u_{AB}$, the unit vector pointing from particle $A$ to $B$ (Fig.~\ref{fig_model}). We assume that the torque exerted by such a link on particle $A$ is 
 \be
 \M^{bend}_{A,l}=k_1 (\t_{A,l} \times \u_{AB} ),
 \label{Mbend}
 \ee
where the microscopic bending rigidity $k_1>0$ is a model parameter. Thus, a stress-free (``straight'') link is pointing in a direction $\t_{A,l}=\u_{AB}$. In general, the torque (\ref{Mbend}) rotates particle $A$ so that $\t_{A,l}$ aligns with $\u_{AB}$ (Figs.~\ref{fig_model}b and c).

We choose Eq. (\ref{Mbend}) due to its simplicity. However, a real mechanical system composed of flexible beams would exert similar torques if particles are much smaller than the length of the interconnecting beams, and beams are softer at their ends hence deformations are localized to the vicinity of the particles. In such cases the preferred link direction $\t_{A,l}$ is the tangent vector of the link $l$ at the surface of particle $A$, and the tangent vector at the midpoint, $\t_{AB,l}$, is well approximated by $\u_{AB}$. 

Torsion of link $l$ is characterized by two normal vectors, $\n_{A,l}$ and $\n_{B,l}$, assigned to each end of the link (Figs.~\ref{fig_model}a and d).  Their specific orientation (normal to the link) is irrelevant, but in the undeformed state $\n_{A,l} = \n_{B,l}$. For small deformations the torque is assumed to be proportional to the torsion  angle, measured as the angle between the normal vectors as:
 \be
 \M^{twist}_{A,l}= k_2(\n_{A,l} \times \n_{B,l})
 \label{Mtorque}
 \ee
where the model parameter $k_2>0$ is the microscopic torsional stiffness. 

We assume, that in a general situation the net torque of link $l$ acting on particle $A$ is a superposition of the torques associated with bending and torsion: 
 \be
 \M_{A,l} = \M^{bend}_{A,l} + \M^{twist}_{A,l} = k_1 (\t_{A,l} \times \u_{AB} ) + k_2(\n_{A,l} \times \n_{B,l}).
 \label{MA}
 \ee
Similarly, for particle $B$ at the other end of the link $l$:
 \be
 \M_{B,l} = k_1 (\t_{B,l} \times \u_{BA} ) + k_2(\n_{B,l} \times \n_{A,l}),
 \label{MB}
 \ee
where $\u_{BA}=-\u_{AB}$.

If a link $l$ exerts forces $\F_{A,l}$ and $\F_{B,l}$ as well as torques $\M_{A,l}$ and
$\M_{B,l}$ at its endpoints (Fig.~\ref{fig_model}c), the link is in mechanical equilibrium if
 \be
 \F_{A,l}+\F_{B,l}=0
 \ee
and
 \be
 \M_{A,l}+\M_{B,l}+(\r_B-\r_A)\times\F_{B,l}=0.
 \label {torq_balance}
 \ee

To determine the force $\F_{A,l}=-\F_{B,l}$,  we introduce a unit vector orthogonal both to $\n_{A,l}$ and $\u_{AB}$ as
 \be
 \n'_{A,l} = \n_{A,l}\times\u_{AB}.
 \ee
We decompose the forces into orthogonal components as
 \be
 \F_{A,l}=F_{A,l}^\parallel \u_{AB} + F_{A,l}^\perp \n_{A,l} + F_{A,l}^{\perp\prime} \n'_{A,l}. 
 \label{Fcompos}
 \ee
Substituting the composition (\ref{Fcompos}) into Eq. (\ref{torq_balance}) yields
 \be
 \M_{A,l}+\M_{B,l}-(\r_B-\r_A)\times(F_{A,l}^\perp\n_{A,l} + F_{A,l}^{\perp\prime}\n'_{A,l})=0.
 \ee
After evaluating the cross products we obtain
 \be
 \M_{A,l}+\M_{B,l}=\vert \r_B-\r_A \vert (F_{A,l}^\perp \n'_{A,l} - F_{A,l}^{\perp\prime}\n_{A,l}),
 \ee
hence 
 \be
 F_{A,l}^\perp={(\M_{A,l}+\M_{B,l})\n'_{A,l} \over \vert \r_B-\r_A\vert}
 \label{Fperp1}
 \ee
and
 \be
 F_{A,l}^{\perp\prime}=-{(\M_{A,l}+\M_{B,l})\n_{A,l} \over \vert \r_B-\r_A\vert}.
 \label{Fperp2}
 \ee
Finally, $F_{A,l}^\parallel$ is determined by Hook's law as
 \be
 F_{A,l}^\parallel = k_3(|\r_B-\r_A|-\ell_l),
 \label{Fpara}
 \ee
where $\ell_l$ is the equilibrium length of link $l$ and $k_3>0$ is the microscopic elastic modulus, the third model parameter. 

Thus, two particles $A$ and $B$ interconnected by link $l$ are
characterized by $\r_A$, $\r_B$, $\t_{A,l}$, $\n_{A,l}$, $\t_{B,l}$, $\n_{B,l}$ and $\ell_0$. Given these quantities, equations (\ref{MA}), (\ref{MB}), (\ref{Fperp1}), (\ref{Fperp2}) and (\ref{Fpara}) allow the calculation of the 9 components of $\F_A=-\F_B$, and of $\M_A$ and $\M_B$.

\subsection{Mechanical equilibrium}
A particle may be attached to multiple links and also be the subject of external forces or torques.
Net forces  
 \be
 \F_i = \F_i^{ext} + \sum_{l \in L_i} \F_{i,l} 
 \ee
and torques
 \be
 \M_i = \M_i^{ext} + \sum_{l \in L_i}\M_{i,l} 
 \ee
are calculated by summation over $L_i$, the set of links associated with particle $i$.  In mechanical equilibrium $\F_i$ and $\M_i$ are zero for each particle $i$. The equilibrium configuration, however, is difficult to obtain directly due to the nonlinear dependence of the forces  on particle positions. Instead, we utilized the following overdamped relaxation process:
 \be
 \dot{ \r_i} = \P_i\F_i 
 \label{relaxF}
 \ee
and 
 \be
 \dot{ \bphi_i} = \Q_i\M_i.
 \label{relaxM}
 \ee
where $\P_i$ and $\Q_i$ are projector matrices to constrain the movement and rotation of node $i$, respectively. For unconstrained particles $\P_i$ and $\Q_i$ are identity matrices.
As the particles rotate, the unit vectors of neutral link direction $\t^{(l)}_i$ and orientation $\n^{(l)}_i$ associated to link $l$ and particle $i$ are updated according to (\ref{rot1}) and (\ref{rot2}).

For a given initial condition, the configuration corresponding to mechanical equilibrium is calculated by solving the coupled ordinary differential equations (\ref{relaxF}) and (\ref{relaxM}) by a fourth order Runge-Kutta method. The relaxation was terminated when the magnitude of the net total force and torque in the system fell below a threshold value.

\subsection{Initial condition, connectivity}
Two dimensional initial conditions were generated by randomly positioning $N$ particles in a square of size $L=\sqrt{N}$. Thus, the spatial scale unit of the simulations is set as the average cell size, $\sim$10 $\mu$m and the simulated 2D cell density is 1 cell/unit area.  In the initial condition we enforced that the distance of two adjacent particles is greater than $d_{min}=0.8$. Particles that are Voronoi neighbors are connected by links when their distance is less than $d_{max}=2$. This rule yields a mean link length of $d_0\approx1.2$.  For a stress-free initial configuration we set the $\t_{i,l}^{(0)}$,  $\n_{i,l}^{(0)}$ vectors as well as the equilibrium link lengths $\ell_l$ so that no internal forces or torques are exerted in the system.

\subsection{Plasticity}
Tissue plasticity is modeled by specific rules that reconfigure the links. As cells can both form new intercellular adhesions and remodel existing ones, in our model the topology of connections changes in time.  In particular, we assume that the lifetime of a link is reduced by tensile forces. For a given link, $l$, the probability of its removal during a short time interval $\Delta t$ follows Bell's rule \cite{Bell78} as
 \be
 p_l\Delta t = A e^{F_l/F^*} \Delta t,
 \label{link_removal}
 \ee 
where $F^*$ is a threshold value, and $A$ is a scaling factor which sets the fragility of the connections.

Two adjacent particles (Voronoi neighbors), $i$ and $j$, can establish new contacts if their distance $d_{i,j}$ is less than $d_{max}$ . During a short time interval $\Delta t$, the probability of this event is a decreasing function of the distance:
 \be
 q_{i,j}\Delta t = B \left(1 - {d_{i,j} \over d_{max} } \right) \Delta t.
 \label{link_insert}
 \ee
The scaling factor $B$ represents the level of cellular protrusive activity devoted to scanning the environment and the ability to form intercellular contacts. 

Simulations are event-driven: using the probability distributions (\ref{link_removal}) and (\ref{link_insert}), we generate the next event $\mu$ and waiting time $\tau$ according to the stochastic Gillespie algorithm \cite{Gillespie77}. The waiting time until the next event is chosen from the distribution 
 \be
 \log P(\tau) = {-\tau\over \sum_l p_l + \sum_{i,j} q_{i,j} }
 \ee
where the sums are evaluated using all possible particle pairs $i,j$ not connected by a link as well as enumerating existing links $l$.

After each event, the system is relaxed into a mechanical equilibrium. A new link is assumed to be stress free immediately after insertion. Cells, however, are expected to maintain a certain area or volume, characterized by the mean stress free link length $d_0$. To reflect this process in our model, the equilibrium length of each link evolves in time according to the stochastic dynamics
 \be
 {d \ell_l \over dt } = C (d_0 - \ell_l) + \xi
 \label{linklength}
 \ee
where $\xi$ is an uncorrelated white noise with variance $\sigma$.

The simulation time is set by the $A$ and $B$ parameters. We set our time unit as $1/B\approx 1$s, the time needed for two adjacent cells to establish a mechanical link. The time needed for a cell-cell contact to mature is $B/C \approx 1$min. We also set the lifetime of an unloaded link to $B/A \approx 1$min, thus two cells pulled away by a force $F^*$  separate in $\sim20$s.

\section{Results}
\subsection{Elastic parameters}
To establish the connection between the macroscopic material parameters such as Young's modulus and the microscopic model parameters $k_1$, $k_2$ and $k_3$, we simulated elastic deformations in response to uniaxial tension, in-plane shear and (three dimensional) plate bending. The simulations started from a stress-free initial condition and the same microscopic parameters were assigned to each particle. During these simulations the connectivity of the particles and the equilibrium link properties do not change, hence the system exhibits a pure elastic behavior. The forces and torques exerted by the links are proportional to the microscopic parameters $k_1$, $k_2$ and $k_3$. Thus, in the simulations we set the scale of forces as $F_0=k_3d_0$, i.e.,
the force required to extend a cell twofold along one direction. This choice of force unit allowed us to perform the simulations with $k_3=1$.

\subsubsection{Uniaxial tension.}
\begin{figure}
\begin{center}
\includegraphics[width=5in]{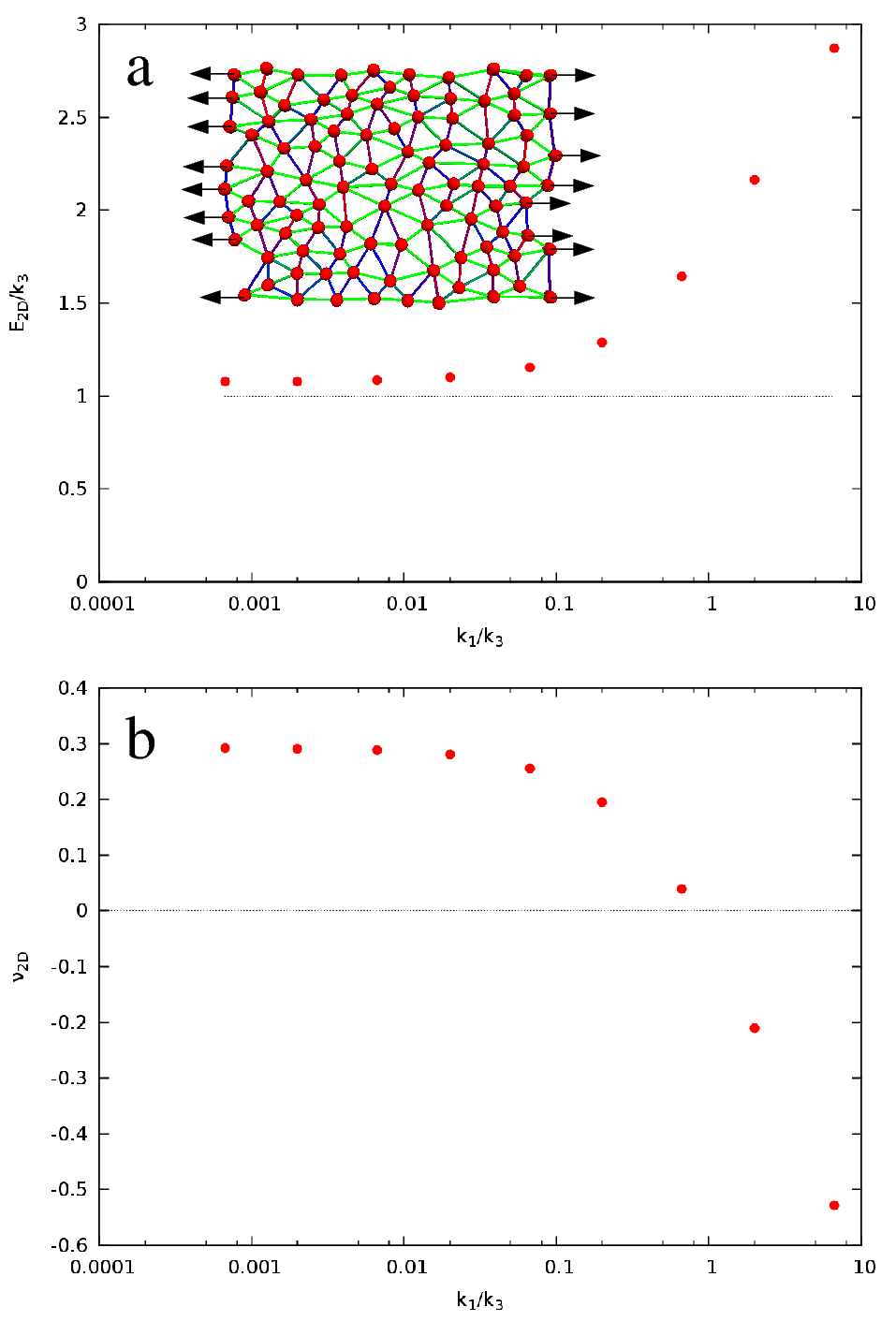}
\caption{\small
Simulations of uniaxial stretch. The dimensionless 2D Young's modulus ($E_{2D}/k_3$; in panel a) and the 2D Poisson's number (b) are plotted as a function of the bending rigidity parameter $k_1/k_3$. The inset in panel (a) depicts a typical configuration of the stretched sample. Black arrows indicate the external forces prescribed during the simulation. The color of the links indicate compression (red), tension (green) or being at the neutral length $d_0$ (blue).
}
\label{fig_stretch}
\end{center}
\end{figure}
We applied external, outward-directed forces on particles that are on the left and right side of the test object (Fig.~\ref{fig_stretch}a, inset). The external force acting on particle $i$ is
 \be
 \F^{ext}_i=\cases{
      -(F_{left},0,0), \mbox{~~for~~} x_i<0.1 L \\
      (F_{right},0,0), \mbox{~~for~~} x_i>0.9 L \\
      (0,0,0), \mbox{~~otherwise.}
   }
 \label{uniax}
 \ee
Parameters $F_{left}$ and $F_{right}$ are chosen in such a way that the net external force is zero and the magnitude of external forces acting on either side is $\cal F$
 \be
 \sum_i \vert \F_i^{ext} \vert = 2{\cal F}.
 \label{extforce}
 \ee

After obtaining mechanical equilibrium, we determined the axial elongation $\Delta L$ and transverse contraction $\Delta W$. The 2D elastic (Young's) modulus was calculated from the engineering stress and strain as
 \be
 E_{2D}={{\cal F}/L \over \Delta L / L } = {{\cal F} \over \Delta L} = 1/a
 \ee
where $a$ is the slope of a linear fit over the $\Delta L$ vs ${\cal F}$ data points, obtained in the range of $0 < {\cal F}/F_0 < 3.2$.
Similarly, the 2D Poisson's ratio is obtained as
 \be
 \nu_{2D} = - {\Delta W \over \Delta L} = b/a,
 \ee
where $b$ is the slope of a linear fit over the $\Delta W$ vs ${\cal F}$ data points.

Since the deformation is planar, interparticle links do not twist. Therefore, $E_{2D}$ and $\nu_{2D}$ do not depend on the value of $k_2$. The bending rigidity parameter $k_1$, however,  can substantially influence both material parameters (Fig.~\ref{fig_stretch}). For $k_1 \ll k_3$, the external forces are mainly balanced by the elongation  of the interparticle links. In this regime thus $E_{2D}\approx k_3$.  The 2D Poisson's ratio is well approximated by $1/[2 tg(\pi/3)]\approx0.29$, the change in the aspect ratio of a triangular lattice when stretched in one direction while keeping the length unchanged for the rest of lattice links. In contrast, for $k_1 \gtrsim k_3$ bending rigidity of the nodes substantially influences the elastic behavior of the system. Interestigly, $\nu_{2D}<0$ for $k_1 \gg k_3$ as link angles are maintained to such an extent that uniaxial stretching will result in an isotropic expansion of the entire mesh. As this regime is biophysically implausible, we restrict the bending rigidity parameter in the 
 \be
 k_1<k_3
 \ee
regime.

\subsubsection{In-plane shear.}
\begin{figure}
\begin{center}
\includegraphics[width=5in]{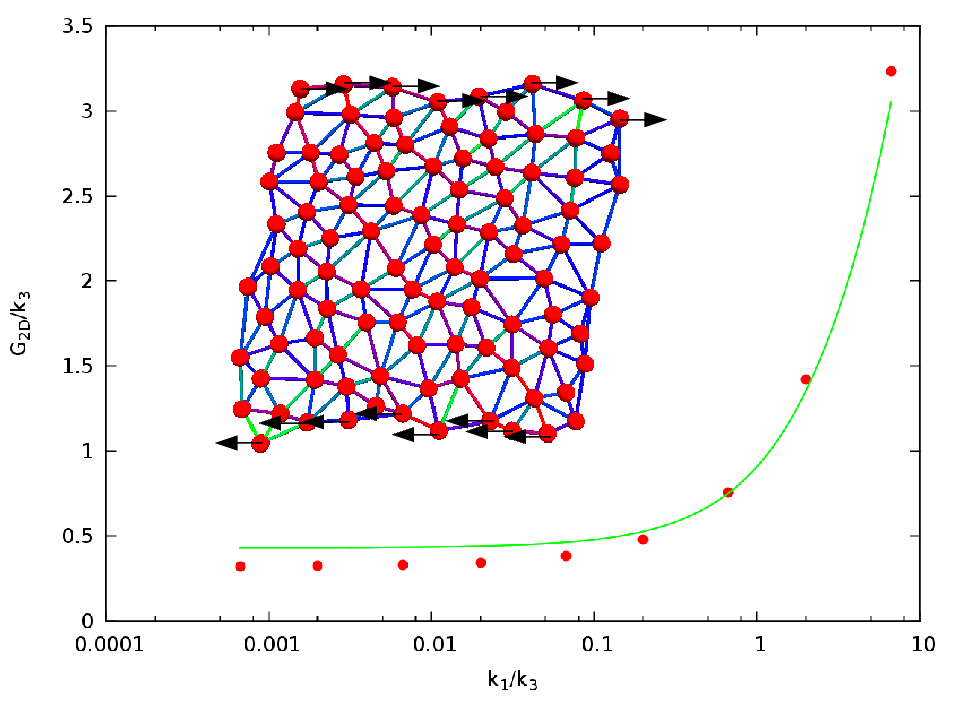}
\caption{\small
The 2D shear modulus, $G_{2D}$, as obtained from simulations.
The green line indicates $G=E/2(1+\nu)$, the relation expected to hold for elastic parameters of a homogenous and isotropic material. The inset depicts a typical configuration of the sheared structure using the color-convention of Fig.~\ref{fig_stretch}.
}
\label{fig_shear}
\end{center}
\end{figure}
In these simulations external shear forces were applied in two layers
as
 \be
 \F^{ext}_i=\cases{
     -(F_{left},0,0), \mbox{~~for~~} y_i<0.1 L \\
     (F_{right},0,0), \mbox{~~for~~} y_i>0.9 L \\
     (0,0,0), \mbox{~~otherwise,}
  }
 \ee
so that the net external force is zero, and the total force is given by (\ref{extforce}). To avoid rotation of the simulated system, particles subjected to external forces were also constrained to move along the x axis.

The shear deformation was quantified using the mean displacements $\Delta x$ of the stripes where forces were acting. The 2D shear modulus was calculated as
 \be
 G_{2D}={{\cal F}/L \over \Delta x / L } = {1\over c} 
 \ee
where $c$ is the slope of a linear fit over the $\Delta x$ vs ${\cal F}$ data points. The obtained shear moduli  are well approximated by 
 \be
 G_{2D}=E_{2D}/2(1+\nu),
 \ee
the relation expected for an isotropic homogeneous linear elastic solid (Fig.~\ref{fig_shear}).

\subsubsection{Bending.}
\begin{figure}
\begin{center}
\includegraphics[width=5in]{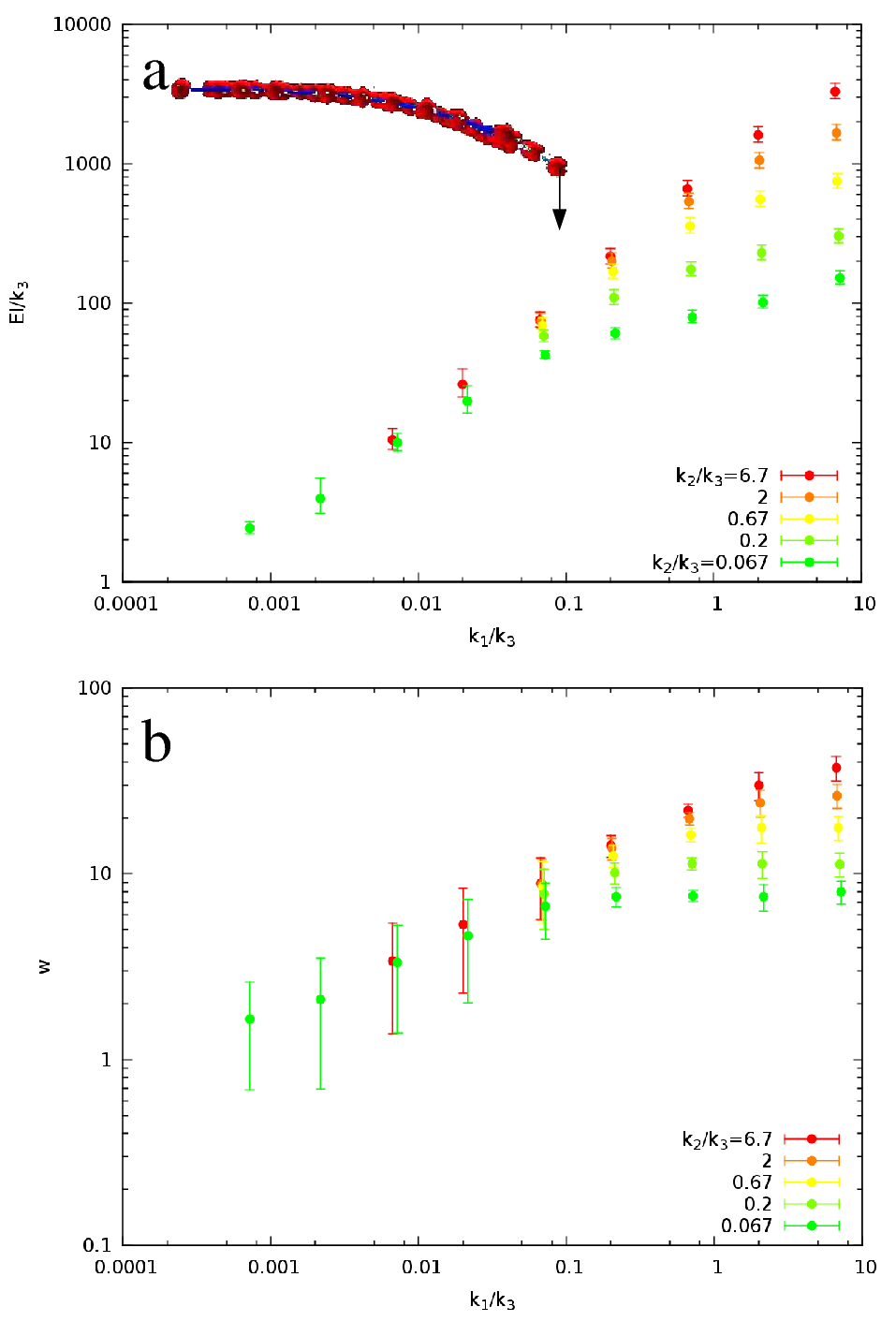}
\caption{\small
Plate bending. a: Macroscopic bending rigidity, $EI$, vs the microscopic bending rigidity $k_1$ for various values of $k_2$ indicated by red to green colors. The inset depicts a typical simulation configuration. b: The thickness $w$ of a plate that exhibits the same bending and Young's moduli as presented in panel (a) and Fig.~\ref{fig_stretch}a.
}
\label{fig_flex}
\end{center}
\end{figure}
To calibrate the bending rigidity of the modeled tissue layer, we simulated a plate immobilized along one side and bent by a perpendicular force exerted at the opposite side.  The loading force was localized at the rightmost 10\% of the particles, acting in a direction perpendicular to the plane of the particles (Fig.~\ref{fig_flex}).

Simulations revealed that the longitudinal cross section of the deflected monolayer is well approximated by a cubic function -- as expected from a beam with finite thickness. Thus, despite the fact that in our model bending rigidity does not  arise through the Euler-Bernoulli/Love-Kirchhoff mechanism, we defined and used the modulus $EI$ to relate the deflection $z$ to the loading force ${\cal F}$ as
 \be
 z=  {\cal F} {L^3 \over 3 EI }.
 \label{flex}
 \ee 
This ``effective'' bending modulus can be tuned over several orders of magnitude, and it is a monotonic, nonlinear function of the microscopic model parameters $k_1$ and $k_2$ (Fig.~\ref{fig_flex}). When torsion of the links is negligible ($k_2 \gg k_1$), the bending modulus $EI$ is proportional to the microscopic bending modulus $k_1$. In contrast, when $k_2 < k_1$ the macroscopic curvature of the simulated sheet is mainly accommodated by torsion of the links, hence in this regime $EI$ depends less on the value of $k_1$.

The bending modulus $EI$ also allows one to associate an effective layer thickness $w$ to a set of microscopic model parameters. For a homogenous elastic material of thickness $w$ and the same lateral size $L$ the second moment of the cross-sectional area is
 \be
 I = {1\over 12} L w^3.
 \ee
If the stretch data shown in Fig.~\ref{fig_stretch} were measured on the same material with a cross section of $wL$, then its (three dimensional) Young's modulus was
 \be
 E = {{\cal F}/Lw \over \Delta L/L} = E_{2D}/w.
 \ee
Using the bending modulus values $EI$ shown in Fig.~\ref{fig_flex}, and substituting $E$ and $I$ into (\ref{flex}) we obtain
 \be
 z=  {\cal F} {4 L^2 \over E_{2D} w^2 }
 \ee
yielding 
 \be
 w=2L \sqrt{ {1\over E_{2D} } {{\cal F} \over z }}.
 \ee
As an example, the parameters $k_1=k_2=10^{-2}k_3$ are consistent with a plate that is composed of columnar cells with an aspect ratio of 1:3 (Fig.~\ref{fig_flex}b).

These results allow further refinement of our force scale $F_0=k_3d_0$ as follows. According to Fig.~\ref{fig_stretch}, the 2D Young's modulus is well approximated by $k_3$ in the $k_1\ll k_3$ regime. The corresponding 3D elastic modulus is then $E=k_3/w$. For the avian epiblast both $E$ and $w$ values are reported in the literature. Using the values of $E_0\approx1$ kPa \cite{Agero10} and $w_0\approx 30 \mu$m \cite{chick_atlas}, we obtain 
 \be
 F_0 = k_3 d_0 \approx E_0 w_0 d_0  = 300 nN
 \ee
for the value of our force unit.

\subsection{Plastic behavior}
Link remodeling allows the simulation of elasto-plastic behavior. As the probability of link removal depends on its load, mechanical stresses are thus accommodated and relaxed in an irreversible process. To characterize macroscopic plastic behavior, we simulated two standard, external force-driven processes: force relaxation within a pre-stressed configuration and creep under uniaxial tension. In both scenarios the initial condition involved external forces that pulled longitudinal links with an average force of $F_0$. 
\subsubsection{Relaxation of mechanical tension.}
\begin{figure}
\begin{center}
\includegraphics[width=5in]{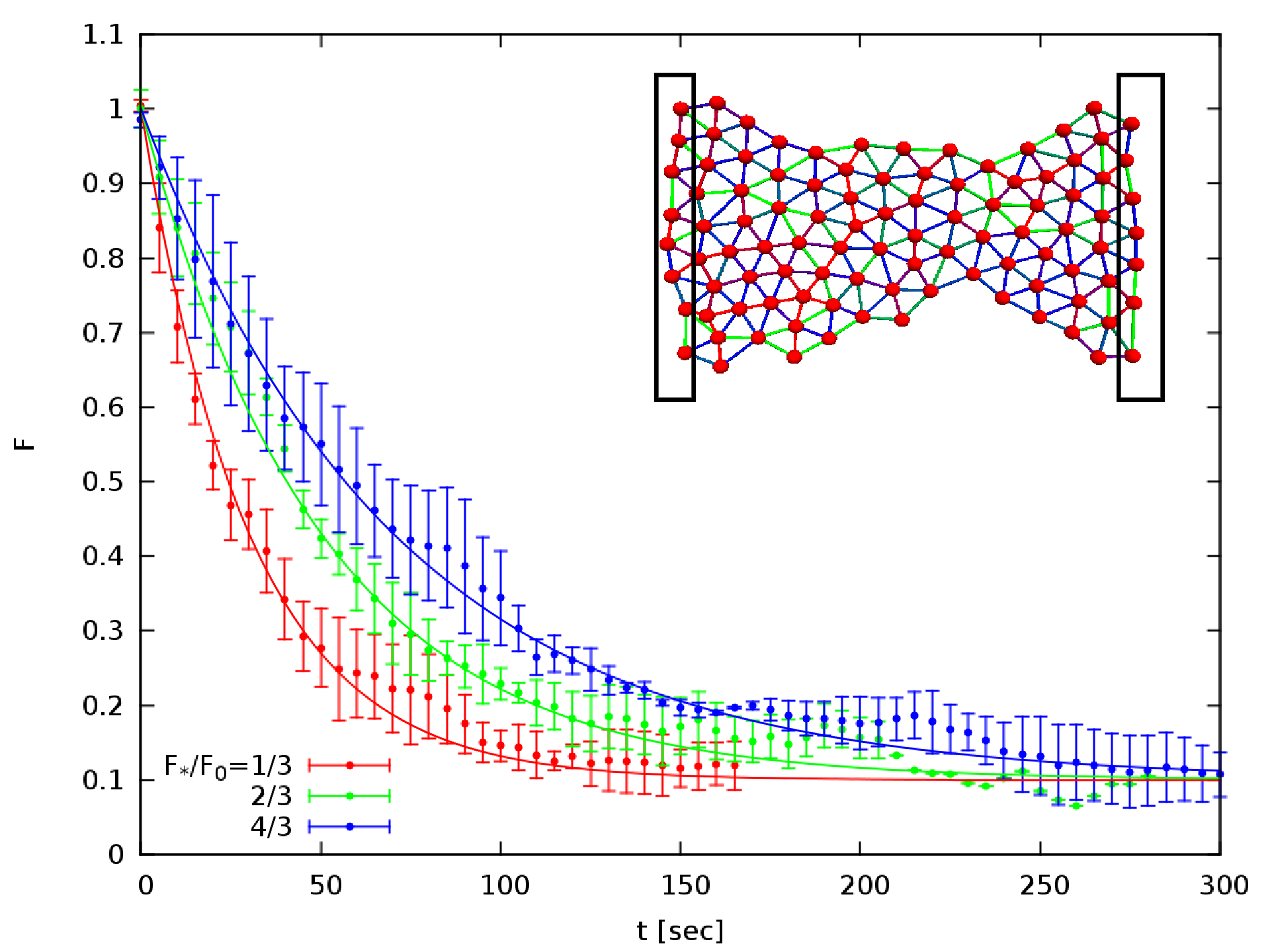}
\caption{\small
The external force required to maintain a pre-set stretch is decreasing in time. A stretched configuration (shown in Fig.~\ref{fig_stretch}) served as the initial condition for the simulations. Particles that had been subjected to external fores were fixed in space (marked by rectangles in the inset). The magnitudes of the external forces required to maintain this constraint are shown normalized to their initial, pre-relaxation values. Data were obtained from stochastic simulations with three choices of the force threshold parameter $F_*$.   Each data point is an average of three independent simulation runs. Error bars indicate standard deviation. The solid curves are exponential functions with characteristic decay times of $\tau=30$s (red), 50s (green) and 70s (blue). The inset depicts a late-stage configuration when only 10\% of the initial external force is needed to maintain the pre-set constrain.
}
\label{fig_relax}
\end{center}
\end{figure}
To investigate how the simulated structure can relax mechanical stresses, first an elastic mechanical equilibrium was obtained in the presence of external uniaxial tensile forces according to (\ref{uniax}). Then, both sides of the stretched sample were fixed in space by replacing the external forces of Eq (\ref{uniax}) by stiff springs as
 \be
 \F_i^{ext} = -k_0 (\r_i - \r_i^{(0)} )
 \label{fix1}
 \ee
where the parameter $k_0=10k_3$ sets the stiffness of the constraint and $\r_i^{(0)}$ denotes the position of particle $i$ at the onset of the plastic relaxation process. This initial condition was used for the plasticity algorithm that generated a series of link removal and insertion events, each followed by updating the configuration to reflect the new mechanical equilibrium. In each configuration, we evaluated the external forces that were needed to maintain the pre-set extension. As Fig.~\ref{fig_relax} demonstrates, the net force exerted at either side decays in time as an exponential function to a positive value (the yield stress), and the characteristic time is approximately proportional to the critical force parameter $F_*$. The magnitude of the characteristic times are in good agreement with the experimental data of \cite{Forgacs98}. During the process particles are rearranged in such a way that the isotropy of link lengths (i.e. cell shape) is restored. The shortening of the longitudinal links was achieved by intercalation: cells in adjacent rows moved into a single row. Hence, in this setting cell intercalation is driven by an external mechanical load. Forces do not diminish completely as the stressed bonds are likely to form again if the mechanical stress within the simulated tissue is too small to move particles further apart.

\subsubsection{Creep under constant tension.}
\begin{figure}
\begin{center}
\includegraphics[width=5in]{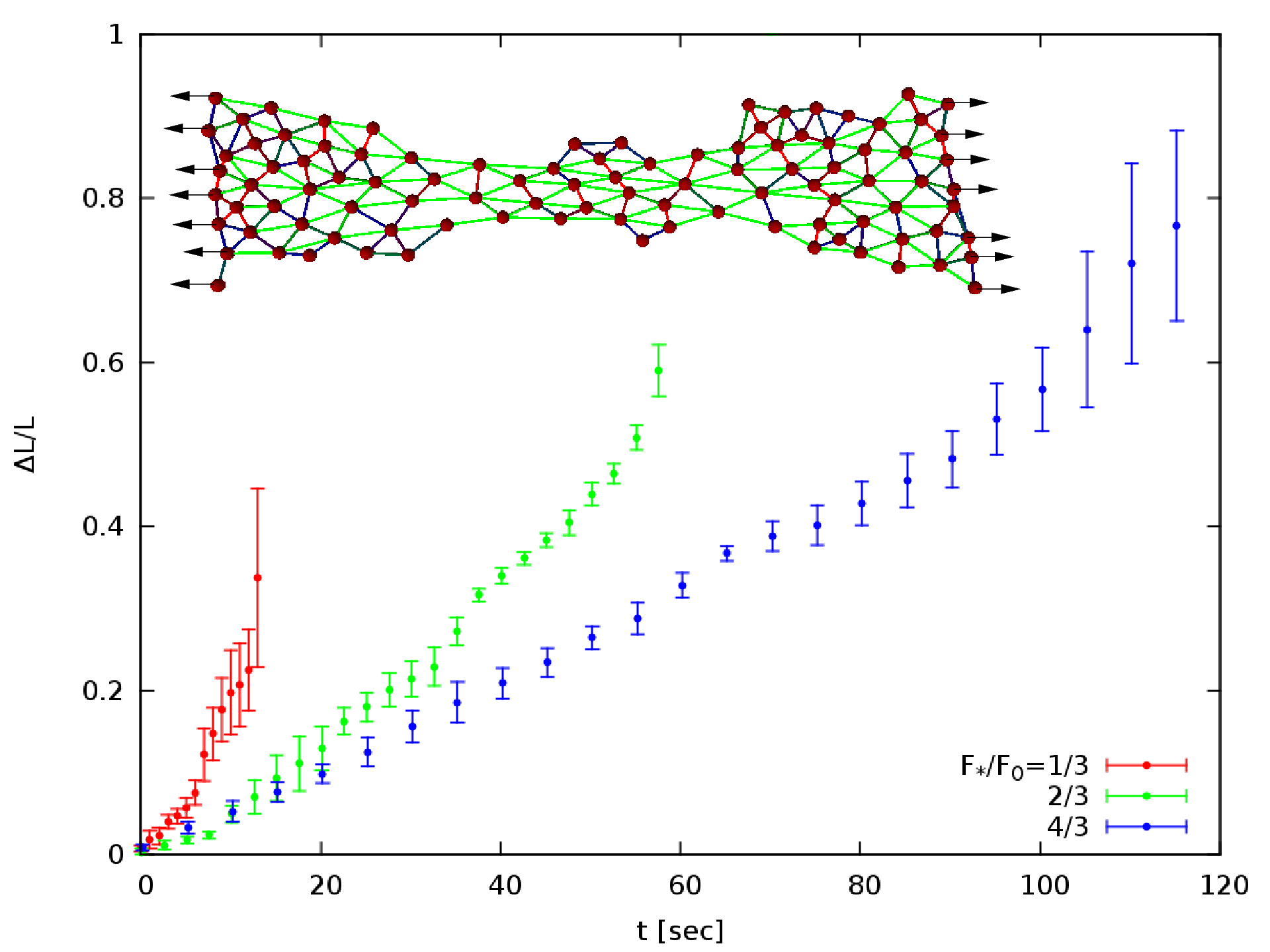}
\caption{\small
Creep during uniaxial tension. The strain of the sample is plotted as a function of time for three force threshold parameters $F_*$, indicated with distinct colors. Each data point is an average of at least three independent simulation runs. Error bars indicate standard deviation.  The inset depicts the configuration after the appearance of necking.
}
\label{fig_creep}
\end{center}
\end{figure}

In a set of complementary simulations the loaded edges were hardened to prevent detachment of the particles. Thus, particles exposed to external forces maintained their orientation by setting $\Q_i$ to null matrices in Eq.~(\ref{relaxM}) and were connected by more stable links (i.e.,  links with increased  $F^*$ parameter). In these simulations the sample gradually extends and narrows (Fig.~\ref{fig_creep}). The initial strain rate is set by the magnitude of the external load and $F^*$. However, as  links are removed faster than new ones are inserted, the strain rate increases with time (necking).

\subsection{Cell movements and active cell adhesion control}
Model particles can be also considered as simulation agents, executing certain prescribed actions in addition to passively responding to the mechanical forces exerted. Such actions may involve the modulation of link parameters such as the equilibrium link length or the neutral link direction. Furthermore,  the probabilities associated with the formation or severance of links may also be controlled. These potential control mechanisms allow the simulation of various cell autonomous behavior and their collective effects.

\subsubsection{Active intercalation.}
\begin{figure}
\begin{center}
\includegraphics[width=5in]{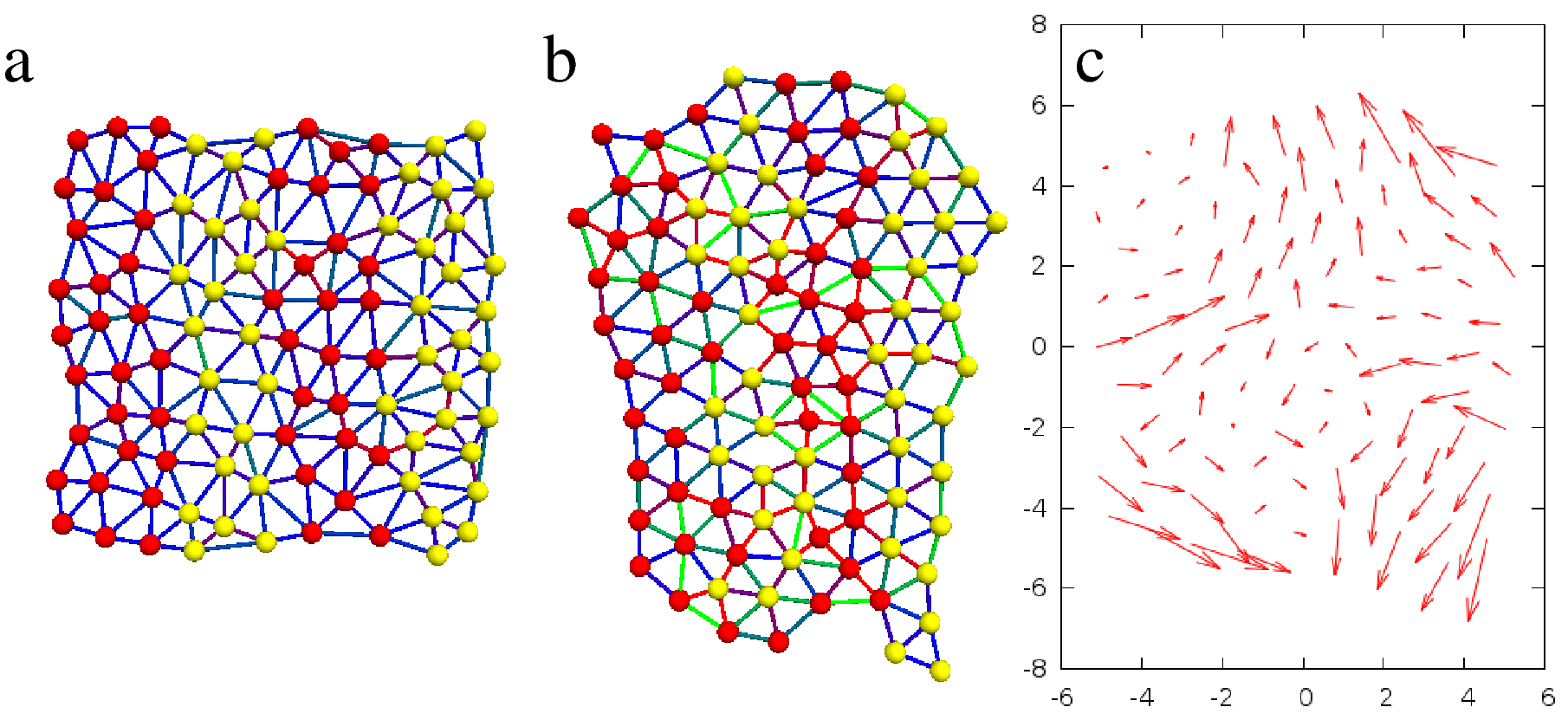}
\caption{\small
Convergent-extension movements driven by active intercalation. a: initial configuration. b: configuration after 300  link insertion or removal events. c: Particle displacements indicate the lateral contraction and axial elongation.
}
\label{fig_convext}
\end{center}
\end{figure}

Anisotropic cell activities were suggested to drive active intercalation movements of early vertebrate embryos \cite{Voiculescu07}. In our model link removal events allow adjacent particles to rearrange their connections. To model active intercalation we assume that cells are more likely to extend processes and establish intercellular contacts along a selected direction ($\p$). Thus Eq.~(\ref{link_insert}) is expanded as
 \be
 q_{i,j}\Delta t = B \left(1 - {d_{i,j} \over d_{max} } \right) 
                  \left(1 - \alpha + \alpha (\u_{i,j}\p)^2 \right) \Delta t,
 \label{link_insert2}
 \ee
where $0\leq \alpha \leq 1$ is a parameter tuning the strength of anisotropy. Similarly, links perpendicular to $\p$ are expected to be less stable, reflected by the modified expression (\ref{link_removal})
 \be
 p_l\Delta t = A e^{F_l/F^*}  
               \left(1 - \alpha(\u_{l}\p)^2 \right) \Delta t.
 \label{link_removal2}
 \ee 
  
Simulations performed with orientation-dependent link probabilities ($\alpha=1$), and with random cell detachments as a driving mechanism, yield both the local cell intercalation and the gradual elongation and lateral contraction (i.e., convergent-extension movements) of the tissue (Fig.~\ref{fig_convext}).

\subsubsection{Autocorrelation functions.}
\begin{figure}
\begin{center}
\includegraphics[width=5in]{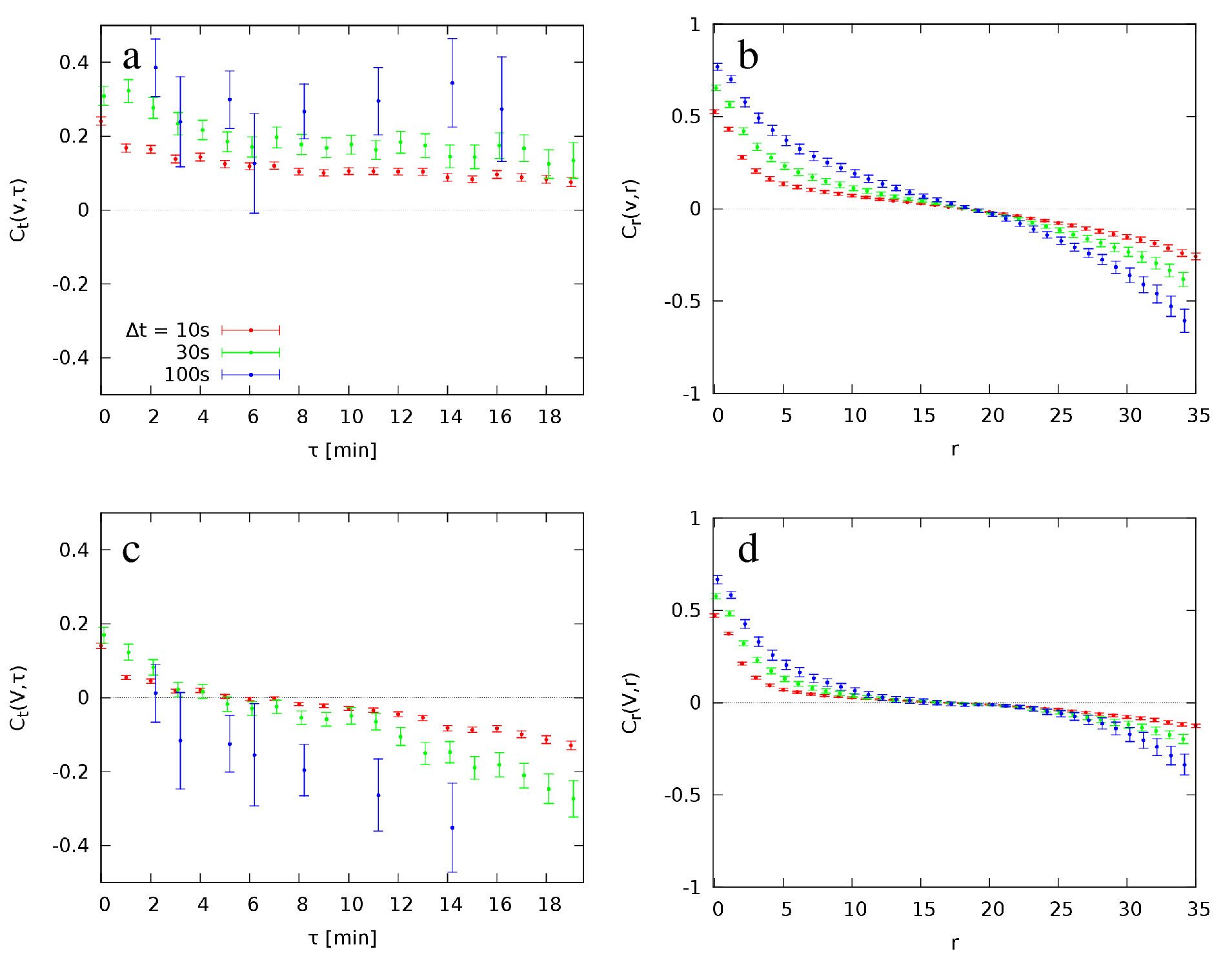}
\caption{\small
Autocorrelation functions of tissue movements obtained in a system of $N=1100$ particles. Temporal (a,c) and spatial (b,d) autocorrelations are plotted for both velocities $v$ (a,b) and velocity fluctuations $V$ (c,d). Velocities are calculated as displacements during a time interval $\Delta t$. Data are shown for three time intervals, $\Delta t =10s$ (red) 30s (green) and 100s (blue). The linear size of the system was $L=33$. 
}
\label{fig_corr}
\end{center}
\end{figure}

The spatial and temporal correlations of tissue movements can be used to characterize both simulations (Fig.~\ref{fig_corr}) and empirical data \cite{Szabo11}.  For an arbitrary quantity $\phi(\x,t)$, temporal autocorrelations are calculated using 
 \be
 C_t(\phi,\tau)={\langle \phi(\x,t') \phi(\x,t) \rangle_{\x,|t'-t|\in B(\tau)} \over \langle \phi^2(\x,t) \rangle_{\x,t} },
 \label{Ct}
 \ee
where $\langle ... \rangle_{\x,t}$ denotes averaging over all possible locations $\x$ and time points $t$. In the nominator of (\ref{Ct}) the time points $t$ and $t'$ are restricted so that their difference falls into a bin $B(x)=[x-b ; x+b]$. Similarly, for spatial autocorrelations we evaluate
 \be
 C_r(\phi,r)={\langle \phi(\x,t) \phi(\x',t) \rangle_{|\x-\x'|\in B(r),t} \over \langle \phi^2(\x,t) \rangle_{\x,t} },
 \label{Cr}
 \ee

Velocities were obtained as
 \be
 \v_i(t) = { \x_i(t+\Delta t) - \x_i(t) \over \Delta t }, 
 \ee
where the $\Delta t$ time interval is a parameter. Particle velocities, driven by link remodeling events and  subsequent relaxation to mechanical equilibrium, exhibit both sustained temporal and long-range spatial correlations (Fig.~\ref{fig_corr}). The latter extend over distances comparable to the size of the simulated system.
Correlations increase for velocities calculated with longer $\Delta t$ values. Hence, instantaneous velocity fields are dominated by mechanical adaptation to random link remodeling events. However, a longer time lag suppresses the noise and the resulting velocity fields are characteristic for the overall tissue movements that are highly correlated both in time and space. Thus, our stochastic simulation rules yield persistent large-scale (multi-cellular) motion patterns.

The presence of a sustained movement pattern motivates to define velocity fluctuations as 
 \be
 \V_i(t)=\v_i(t) - \overline{\v}_i
 \ee
where $\overline{\v}_i = \langle \v_i(t)\rangle_t$ is the sustained drift velocity of particle $i$. The velocity fluctuations, mainly adjustments due to changes in connectivity, do not show long range temporal correlations (Fig.~\ref{fig_corr}c). The finite correlation time $\sim 2$min reflects the link length adjustment rule (\ref{linklength}). Velocity fluctuations, however, continue to exhibit long-range spatial correlations (Fig.~\ref{fig_corr}d), due to the strong mechanical coupling within the system.

\section{Discussion}

\subsection{Tissue plasticity}
\begin{figure}
\begin{center}
\includegraphics[width=5in]{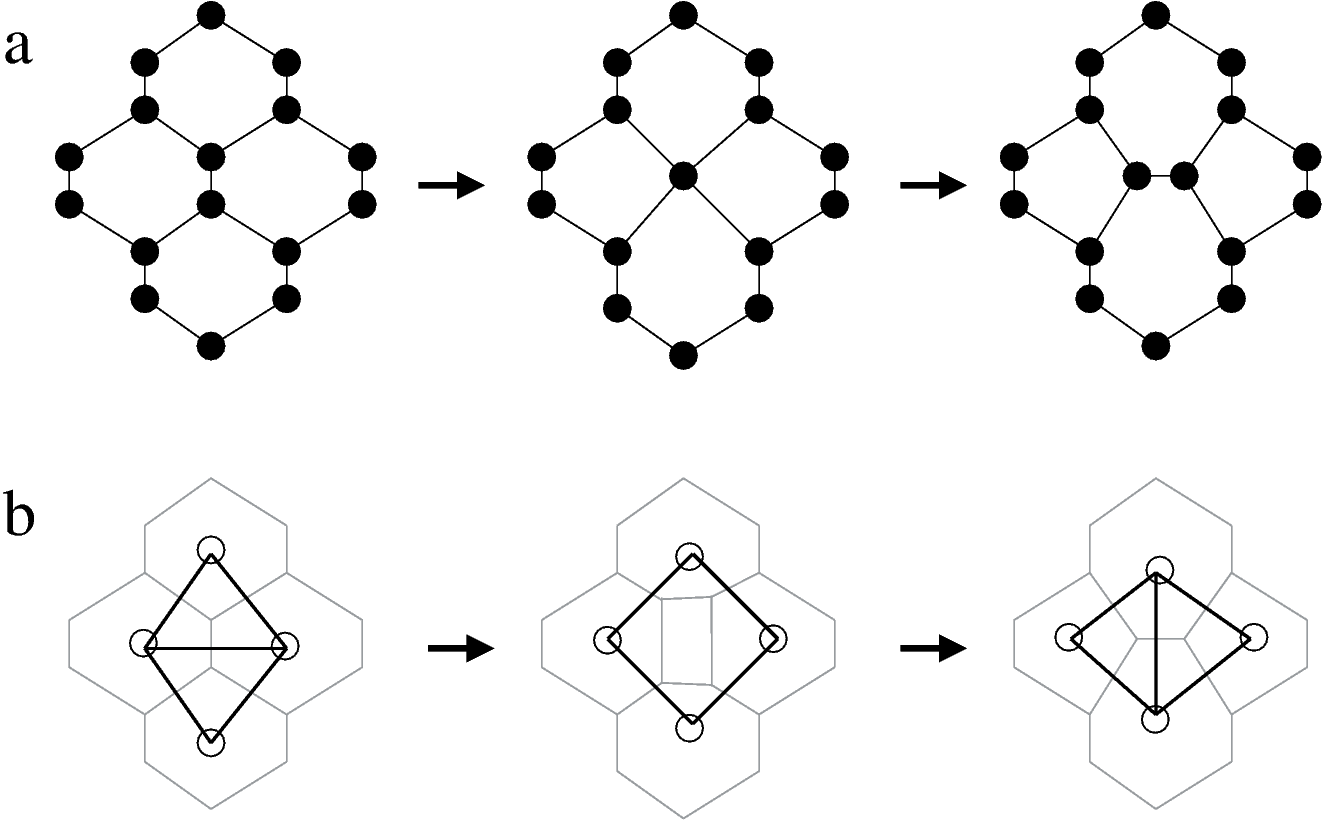}
\caption{\small
Cell neighbor exchange in a T1 transition. a: In the vertex model a polygonal boundary is eliminated, then a new boundary segment forms between cells that were previously non-adjacent. Vertices are represented by filled circles. In the model proposed here a stretched connection between two adjacent cells is removed. As the link was load-bearing, a mechanical equilibrium yields a new configuration where the previously adjacent cells move away, while their neighbors move closer in a direction perpendicular to the direction of the removed link. The central ``hole'' in the second configuration indicates a soft patch into which adjacent cells can protrude.  Finally, a new link can connect previously non-adjacent particles. Particles are represented by open circles, black segments indicate the links between particles. Gray polygons are cell shapes that are not resolved in the model. 
}
\label{fig_t1}
\end{center}
\end{figure}

The plastic behavior of cell aggregates was well studied in a series of experiments where aggegates were compressed in an apparatus where both the compression force as well as the shape of the deformed sample could be precisely monitored \cite{Forgacs98}.  Under these circumstances, a short compression elicits an elastic (reversible) deformation. In this elastic regime the deformation is homogenous within the aggregate: individual cells are also compressed along one direction.  In contrast, when the compression is maintained for several hours, the force needed to maintain the deformation decreases in time. This decrease is close to exponential, and the behavior is plastic in the sense that after removing the compression force, the aggregate remains in the new equilibrium shape for several hours. Confocal microscopy revealed that cells within the aggregate regained their isotropic shapes  -- hence they remodeled their intercellular connections. These empirical findings, including the time scale of the plastic relaxation process, are well reproduced by our model (Fig.~\ref{fig_relax}).

Exponential relaxation of shear stress, characteristic for the Maxwell fluid, has been proposed to model embryonic tissues. Such behavior can arise by a sufficient number of cell divisions or apoptoses within the tissue \cite{Ranft10}, or by a mechanical load-dependent remodeling of intercellular connections \cite{Preziosi10, Preziosi11}. In our simulations, in accord with the continuum theory by Preziosi et al \cite{Preziosi10, Preziosi11}, the stress does not diminish completely. We attribute this effect to an inherent granularity within the model. A stretched link is replaced in a T1 transition (Fig.~\ref{fig_t1}) by the following sequence of events: 1) a stretched link is removed.  2) The distance between the previously interconnected particles is increased in the new mechanical equilibrium. 3) The resulting ``gap'' (or soft patch) is filled in by a new link connecting cells in the orthogonal direction. This last step occurs only if the local deformation in step 2 alters distances to an extent that the two particles that were interconnected by the removed link do not remain Voronoi neighbors. Thus, in our model a yield stress, below which the tissue response is elastic, arises naturally.

\subsection{Intercellular forces}
Bell's rule of force-mediated bond dissociation (\ref{link_removal}) and the corresponding exponential lifetime of adhesion links \cite{Bell78} is supported by dynamic force microscopy experiments \cite{Zhang04}. The reported  studies indicated a single molecule rupture force (corresponding to $F^*$ for a single molecule) around 10 pN. As a cell may display $10^5$ adhesion molecules on its membrane \cite{Foty04},  two adjacent cells may be linked by $10^4$ adhesion molecules. Thus, the rupture force needed to separate two cells is 100 nN if we assume that separation breaks each adhesion bond between the two cells. When cells are pulled apart slowly, a much lower force is sufficient as adhesion molecules can spontaneously unbind: for example, embryonic zebrafish cells can be separated by 10 nN forces \cite{Maitre12}. In our model the separation of a strained intercellular contact is assumed to be instantaneous (not resolved by the dynamics), we estimate $F^*\approx 100nN$, which in our force units translates into $F^* \approx F_0/3$, a value used for simulations in Figs.~\ref{fig_relax}-\ref{fig_convext}.

\subsection{Tissue movements}
The emergence of tissue movements from the collective action of its constituent cells is one of the central questions of developmental biology. Imaging studies established that during later stages of development involving a well-crosslinked matrix the cells and their immediate extracellular matrix surroundings move as a composite material \cite{Benazeraf10,Aleksandrova12}, while during earlier stages individual cell motility is superimposed on a larger scale tissue movement \cite{Zamir06,Zamir08}. Hence the body plan of early amniote embryos do not appear to be established by ``conventional'' cell motility -- i.e., cells migrating on an external substrate to pre-defined positions following environmental cues. Instead, germ layers and the entire embryo morphology are molded to a large extent by cell-exerted mechanical forces (stresses) and their controlled dissipation/relaxation.

Anisotropic cell behavior has been proposed previously to explain convergent-extension movements \cite{Zajac00, Zajac03, Rauzi08, Vasiev10, Sandersius11c}. Here we demonstrated that anisotropic cell activity can be formulated within our proposed model, and the simulations yield velocity autocorrelations comparable with those reported for avian embryos \cite{Szabo11}. In particular, particle image velocimetry revealed that the displacements of morphogenetic tissue movements are smooth in space and tissue movements are correlated even at locations separated by several hundred micrometers, comparable to the size of the embryo. Velocity vectors, however, strongly fluctuate in time. The autocorrelation time of the velocity fluctuations was reported to be less than a minute. Our model suggest that fluctuations with a short correlation time can be generated by sudden cellular detachment/reattachment events occurring at random positions -- the driving mechanism of tissue movements.  The long correlation length is consistent with the idea that the tissue is in mechanical equilibrium, therefore a local change in cell traction is expected to immediately alter tissue deformations elsewhere.

In summary, we demonstrated that the proposed model yields a realistic elasto-plastic behavior with exponential relaxation of tensile stresses above an intrinsic threshold value. Due to the simplicity of the model, microscopic parameter values can be inferred from tissue thickness, its macroscopic elastic modulus and Poisson's number and the magnitude and dynamics of intercellular adhesion forces. The proposed stochastic simulation of cell activities  gives rise to fluctuating tissue movements, which exhibit the same autocorrelation properties as the empirically obtained data. This model, therefore, can serve as a mechanically correct basis for cell-resolved or agent-based future studies focusing on tissue morphogenesis.

\section*{Acknowledgements}
This work was supported by the NIH (grant HL085694 to Dr Brenda Rongish), the
Hungarian Development Agency (KTIA AIK 12-1-2012-0041), the Hungarian Research
Fund (OTKA K72664) and the G. Harold \& Leila Y. Mathers Charitable Foundation.
We are most grateful for Drs Sandra Rugonyi, Brenda Rongish and Charles Little
for fruitful discussions and comments on the manuscript.



\section*{References}


\begin{thebibliography}{10}

\bibitem{Preziosi10}
L.~Preziosi, D.~Ambrosi, and C.~Verdier.
\newblock An elasto-visco-plastic model of cell aggregates.
\newblock {\em J Theor Biol}, 262(1):35--47, Jan 2010.

\bibitem{Preziosi11}
L~Preziosi and Guido Vitale.
\newblock A multiphase model of tumor and tissue growth including cell adhesion
  and plastic reorganization.
\newblock {\em Mathematical Models and Methods in Applied Sciences},
  21(09):1901--1932, September 2011.

\bibitem{Ranft10}
Jonas Ranft, Markus Basan, Jens Elgeti, Jean-Francois Joanny, Jacques Prost,
  and Frank JÃ¼licher.
\newblock Fluidization of tissues by cell division and apoptosis.
\newblock {\em Proc Natl Acad Sci U S A}, 107(49):20863--20868, Dec 2010.

\bibitem{Giverso12}
C.~Giverso and L.~Preziosi.
\newblock Modelling the compression and reorganization of cell aggregates.
\newblock {\em Math Med Biol}, 29(2):181--204, Jun 2012.

\bibitem{Fung93}
Y.~C. Fung.
\newblock {\em Biomechanics: mechanical properties of living tissues}.
\newblock Springer-Verlag, New York, 1993.

\bibitem{Stolarska09}
Magdalena~A Stolarska, Yangjin Kim, and Hans~G Othmer.
\newblock Multi-scale models of cell and tissue dynamics.
\newblock {\em Philos Trans A Math Phys Eng Sci}, 367(1902):3525--3553, Sep
  2009.

\bibitem{Taber09}
Larry~A Taber.
\newblock Towards a unified theory for morphomechanics.
\newblock {\em Philos Transact A Math Phys Eng Sci}, 367(1902):3555--3583, Sep
  2009.

\bibitem{Varner10}
Victor~D Varner, Dmitry~A Voronov, and Larry~A Taber.
\newblock Mechanics of head fold formation: investigating tissue-level forces
  during early development.
\newblock {\em Development}, 137(22):3801--3811, Nov 2010.

\bibitem{Graner92}
Graner and Glazier.
\newblock Simulation of biological cell sorting using a two-dimensional
  extended potts model.
\newblock {\em Phys Rev Lett}, 69(13):2013--2016, Sep 1992.

\bibitem{Izaguirre04}
J.~A. Izaguirre, R.~Chaturvedi, C.~Huang, T.~Cickovski, J.~Coffland, G.~Thomas,
  G.~Forgacs, M.~Alber, G.~Hentschel, S.~A. Newman, and J.~A. Glazier.
\newblock Compucell, a multi-model framework for simulation of morphogenesis.
\newblock {\em Bioinformatics}, 20(7):1129--1137, 2004.

\bibitem{Newman05}
T.J. Newman.
\newblock Modeling multicellular systems using subcellular elements.
\newblock {\em Math. Biosci. Eng.}, 2:611--622, 2005.

\bibitem{Sandersius11a}
Sebastian~A Sandersius, Manli Chuai, Cornelis~J Weijer, and Timothy~J Newman.
\newblock Correlating cell behavior with tissue topology in embryonic
  epithelia.
\newblock {\em PLoS One}, 6(4):e18081, 2011.

\bibitem{Forgacs98}
G.~Forgacs, R.~A. Foty, Y.~Shafrir, and M.~S. Steinberg.
\newblock Viscoelastic properties of living embryonic tissues: a quantitative
  study.
\newblock {\em Biophys J}, 74(5):2227--2234, 1998.

\bibitem{Zajac00}
M~Zajac, G~L Jones, and J~A Glazier.
\newblock Model of convergent extension in animal morphogenesis.
\newblock {\em Phys. Rev. Lett.}, 85:2022--5, 2000.

\bibitem{Vasiev10}
Bakhtier Vasiev, Ariel Balter, Mark Chaplain, James~A Glazier, and Cornelis~J
  Weijer.
\newblock Modeling gastrulation in the chick embryo: formation of the primitive
  streak.
\newblock {\em PLoS One}, 5(5):e10571, 2010.

\bibitem{Sandersius11c}
S.~A. Sandersius, M.~Chuai, C.~J. Weijer, and T.~J. Newman.
\newblock A 'chemotactic dipole' mechanism for large-scale vortex motion during
  primitive streak formation in the chick embryo.
\newblock {\em Phys Biol}, 8(4):045008, Aug 2011.

\bibitem{Szabo11}
A.~Szab\'o, P.~A. Rupp, B.~J. Rongish, C.~D. Little, and A.~Czir\'ok.
\newblock Extracellular matrix fluctuations during early embryogenesis.
\newblock {\em Phys Biol}, 8(4):045006, Aug 2011.

\bibitem{Honda80}
H.~Honda and G.~Eguchi.
\newblock How much does the cell boundary contract in a monolayered cell sheet?
\newblock {\em J Theor Biol}, 84(3):575--588, Jun 1980.

\bibitem{Fletcher14}
Alexander~G Fletcher, Miriam Osterfield, Ruth~E Baker, and Stanislav~Y
  Shvartsman.
\newblock Vertex models of epithelial morphogenesis.
\newblock {\em Biophys J}, 106(11):2291--2304, Jun 2014.

\bibitem{Farhadifar07}
Reza Farhadifar, Jens-Christian Röper, Benoit Aigouy, Suzanne Eaton, and Frank
  Jülicher.
\newblock The influence of cell mechanics, cell-cell interactions, and
  proliferation on epithelial packing.
\newblock {\em Curr Biol}, 17(24):2095--2104, Dec 2007.

\bibitem{Rauzi08}
Matteo Rauzi, Pascale Verant, Thomas Lecuit, and Pierre-François Lenne.
\newblock Nature and anisotropy of cortical forces orienting drosophila tissue
  morphogenesis.
\newblock {\em Nat Cell Biol}, 10(12):1401--1410, Dec 2008.

\bibitem{Maitre12}
Jean-LÃ©on Maitre, HÃ©lÃ¨ne Berthoumieux, Simon Frederik~Gabriel Krens,
  Guillaume Salbreux, Frank JÃ¼licher, Ewa Paluch, and Carl-Philipp Heisenberg.
\newblock Adhesion functions in cell sorting by mechanically coupling the
  cortices of adhering cells.
\newblock {\em Science}, 338(6104):253--256, Oct 2012.

\bibitem{Honda04}
Hisao Honda, Masaharu Tanemura, and Tatsuzo Nagai.
\newblock A three-dimensional vertex dynamics cell model of space-filling
  polyhedra simulating cell behavior in a cell aggregate.
\newblock {\em J Theor Biol}, 226(4):439--453, Feb 2004.

\bibitem{Drasdo00}
D.~Drasdo and G.~Forgacs.
\newblock Modeling the interplay of generic and genetic mechanisms in cleavage,
  blastulation, and gastrulation.
\newblock {\em Dev Dyn}, 219(2):182--191, Oct 2000.

\bibitem{RamisConde08}
Ignacio Ramis-Conde, Dirk Drasdo, Alexander R~A Anderson, and Mark A~J
  Chaplain.
\newblock Modeling the influence of the e-cadherin-beta-catenin pathway in
  cancer cell invasion: a multiscale approach.
\newblock {\em Biophys J}, 95(1):155--165, Jul 2008.

\bibitem{Bell78}
G.~I. Bell.
\newblock Models for the specific adhesion of cells to cells.
\newblock {\em Science}, 200(4342):618--627, May 1978.

\bibitem{Gillespie77}
Daniel~T. Gillespie.
\newblock Exact stochastic simulation of coupled chemical reactions.
\newblock {\em The Journal of Physical Chemistry}, 81(25):2340--2361, 1977.

\bibitem{Agero10}
Ubirajara Agero, James~A Glazier, and Michael Hosek.
\newblock Bulk elastic properties of chicken embryos during somitogenesis.
\newblock {\em Biomed Eng Online}, 9:19, 2010.

\bibitem{chick_atlas}
R.~Bellairs and M.~Osmond.
\newblock {\em The atlas of chick development}.
\newblock Academic Press, San Diego, CA, 1998.

\bibitem{Voiculescu07}
Octavian Voiculescu, Federica Bertocchini, Lewis Wolpert, Ray~E Keller, and
  Claudio~D Stern.
\newblock The amniote primitive streak is defined by epithelial cell
  intercalation before gastrulation.
\newblock {\em Nature}, 449(7165):1049--1052, Oct 2007.

\bibitem{Zhang04}
Xiaohui Zhang, Susan~E Craig, Hishani Kirby, Martin~J Humphries, and Vincent~T
  Moy.
\newblock Molecular basis for the dynamic strength of the integrin
  alpha4beta1/vcam-1 interaction.
\newblock {\em Biophys J}, 87(5):3470--3478, Nov 2004.

\bibitem{Foty04}
Ramsey~A Foty and Malcolm~S Steinberg.
\newblock Cadherin-mediated cell-cell adhesion and tissue segregation in
  relation to malignancy.
\newblock {\em Int J Dev Biol}, 48(5-6):397--409, 2004.

\bibitem{Benazeraf10}
Bertrand Benazeraf, Paul Francois, Ruth~E Baker, Nicolas Denans, Charles~D
  Little, and Olivier Pourquie.
\newblock A random cell motility gradient downstream of fgf controls elongation
  of an amniote embryo.
\newblock {\em Nature}, 466(7303):248--252, Jul 2010.

\bibitem{Aleksandrova12}
Anastasiia Aleksandrova, Andras Czirok, Andras Szabo, Michael~B Filla,
  M.~Julius Hossain, Paul~F Whelan, Rusty Lansford, and Brenda~J Rongish.
\newblock Convective tissue movements play a major role in avian endocardial
  morphogenesis.
\newblock {\em Dev Biol}, 363(2):348--361, Mar 2012.

\bibitem{Zamir06}
Evan~A Zamir, Andras Czirok, Cheng Cui, Charles~D Little, and Brenda~J Rongish.
\newblock Mesodermal cell displacements during avian gastrulation are due to
  both individual cell-autonomous and convective tissue movements.
\newblock {\em Proc Natl Acad Sci U S A}, 103(52):19806--19811, Dec 2006.

\bibitem{Zamir08}
Evan~A Zamir, Brenda~J Rongish, and Charles~D Little.
\newblock The ecm moves during primitive streak formation--computation of ecm
  versus cellular motion.
\newblock {\em PLoS Biol}, 6(10):e247, Oct 2008.

\bibitem{Zajac03}
Mark Zajac, Gerald~L Jones, and James~A Glazier.
\newblock Simulating convergent extension by way of anisotropic differential
  adhesion.
\newblock {\em J Theor Biol}, 222(2):247--259, 2003.

\end{thebibliography}
\end{document}